\documentclass[useAMS,usenatbib]{mn2e}
\usepackage{longtable}
\usepackage{graphicx}
\usepackage{lscape}
\usepackage{rotating}
\newcommand{\mnras}{MNRAS}
\newcommand{\aj}{AJ}
\newcommand{\apj}{ApJ}
\newcommand{\apjs}{ApJ}
\newcommand{\apjl}{ApJ}
\newcommand{\aap}{A\&A}

\newcommand{\pasp}{PASP}

\newcommand{\Mwd}{\mbox{$M_\mathrm{wd}$}}

\newcommand{\Rwd}{\mbox{$R_\mathrm{wd}$}}

\newcommand{\Twd}{\mbox{$T_\mathrm{wd}$}}

\newcommand{\Msun}{\mbox{$\mathrm{M}_{\odot}$}}
\newcommand{\Rsun}{\mbox{$R_{\odot}$}}

\newcommand{\Teff}{\mbox{$T_{\mathrm{eff}}$}}

\newcommand{\Porb}{\mbox{$P_{\mathrm{orb}}$}}

\newcommand{\Lines}[3]{\Ion{#1}{#2}\,$\lambda\lambda$\,#3}
\newcommand{\Ion}[2]{#1{\,\scriptsize #2}}

\newcommand{\kms}{\mbox{$\mathrm{km\,s^{-1}}$}}

\title[The SDSS WDMS  binary catalogue] {Post-common envelope binaries
  from  SDSS  -  XIV.   The  DR\,7 white  dwarf-main  sequence  binary
  catalogue}

\author[A.    Rebassa-Mansergas  et   al.]{A.   Rebassa-Mansergas$^1$,
  A.      Nebot      G\'omez-Mor\'an$^2$,     M.R.      Schreiber$^1$,
  B.T. G\"ansicke$^3$, \newauthor A. Schwope$^4$, J.Gallardo$^5$, 
  D. Koester$^6$\\
$^{1}$ Departamento de F\'\i sica y Astronom\'\i a, Universidad de Valpara\'\i so, 
Avenida Gran Bretana 1111, Valpara\'\i so, Chile \\
$^{2}$ Universit\'e de Strasbourg, CNRS, UMR7550, Observatoire Astronomique  
de Strasbourg, 11 Rue de l'Universit\'e, F-67000 Strasbourg, France\\
$^{3}$ Department of Physics, University of Warwick, Coventry CV4 7AL, UK \\
$^{4}$ Astrophysikalisches Inst. Potsdam, An der Sternwarte 16, 14482,
Potsdam, Germany\\
$^{5}$Departamento de Astronom\'\i a, Universidad de Chile, Casilla 36-D,
Santiago, Chile\\
$^{6}$ Institut f\"ur Theoretische Physik und Astrophysik, University of Kiel,
24098 Kiel, Germany\\
}

\begin{document}
\date{Accepted 2011. Received 2011; in original form 2011}
\pagerange{\pageref{firstpage}--\pageref{lastpage}} \pubyear{2011}
\maketitle

\begin{abstract}
We present  an updated version  of the spectroscopic  white dwarf-main
sequence  (WDMS) binary catalogue  from the  Sloan Digital  Sky Survey
(SDSS).  395  new  systems  are  serendipitous  discoveries  from  the
spectroscopic  SDSS\,I/II Legacy targets.  As part  of SEGUE,  we have
carried  out a  dedicated and  efficient  (64 per  cent success  rate)
search for WDMS  binaries with a strong contribution  of the companion
star, which were underrepresented by all previous surveys, identifying
251  additional systems. In  total, our  catalogue contains  2248 WDMS
binaries, and includes, where available, magnitudes from the GALEX All
Sky Survey in  the ultraviolet and from the  UKIRT Infrared Sky Survey
(UKIDSS) in  the near-infrared. We  also provide radial  velocities of
the  companion stars, measured  from the  SDSS spectroscopy  using the
\Lines{Na}{I}{8183.27,8194.81} absorption doublet and/or the H$\alpha$
emission.    Using    an    updated    version   of    our    spectral
decomposition/fitting  technique we  determine/update the  white dwarf
effective temperatures,  surface gravities and masses, as  well as the
spectral   type    of   the    companion   stars   for    the   entire
catalogue.   Comparing  the   distributions  of   white   dwarf  mass,
temperature, and  companion spectral type,  we confirm that  our SEGUE
survey project  has been successful in identifying  WDMS binaries with
cooler  and more  massive white  dwarfs, as  well as  earlier spectral
types  than found previously.  Finally, we  have developed  a publicly
available interactive  on-line data  base for spectroscopic  SDSS WDMS
binaries   containing  all   available   stellar  parameters,   radial
velocities and magnitudes which we briefly describe.

\end{abstract}


\label{firstpage}

\section{Introduction}
\label{s-intro}

Products  of common envelope  evolution play  important roles  in many
areas  of  modern astronomy,  e.g.   stellar  black  hole binaries  as
laboratories  for general  relativity,  double-degenerate white  dwarf
binaries  as   potential  progenitors   of  type  Ia   supernovae,  or
double-degenerate neutron star binaries  as progenitors of short gamma
ray bursts. The  fundamental concept of the formation  of such systems
is well  established: once  the more massive  star in  a main-sequence
binary evolves into  a red giant, unstable mass  transfer is initiated
at a rate that exceeds the  Eddington limit of the companion star, and
leads to the formation of a  common envelope around the giant core and
its  less   massive  main  sequence   companion  \citep{paczynski76-1,
  webbink84-1, iben+tutukov86-1, iben+livio93-1}.  Drag forces between
the binary components and the material  of the envelope lead then to a
dramatic  decrease of the  binary separation,  and the  orbital energy
released  due to  the shrinkage  of  the orbit  eventually expels  the
envelope exposing a post common-envelope binary (PCEB).

PCEBs  continue to  evolve  to even  shorter  orbital periods  through
angular momentum  loss given by magnetic  braking and/or gravitational
wave  emission,  and may  either  undergo  a  second common  envelope,
leading to double-degenerate PCEBs, or enter a semi-detached state and
appear as  cataclysmic variables,  super-soft X-ray sources,  or X-ray
binaries.  Because  of the complex physical processes,  and wide range
of physical and time scales involved in the common envelope evolution,
our  understanding of  this important  phase is  still very  poor, and
severely under-constrained by observations.

Among the variety of  PCEBs, white dwarf-main sequence (WDMS) binaries
are   intrinsically  the  most   common,  and   structurally  simplest
population, and hold a strong promise to provide crucial observational
input  that is  necessary for  improving  the theory  of close  binary
evolution \citep{schreiber+gaensicke03-1}.  WDMS binaries descend from
main sequence binaries where the  primary has a mass $\la10$\Msun.  In
the  majority of cases  ($\sim$3/4) the  initial main  sequence binary
separation is large enough for  the white dwarf precursor to evolve in
the same  way as  a single star  \citep{dekool92-1, willems+kolb04-1},
and  consequently the orbital  period of  these systems  will increase
because of the mass loss of the primary. In the remaining $\sim$1/4 of
the  cases,  the  system  enters  a  common  envelope,  leading  to  a
drastically shorter orbital period. The orbital period distribution of
the entire  population of  WDMS binaries is  therefore expected  to be
strongly  bi-modal,  with  the  short  orbital  period  PCEBs  clearly
separated  from the  long orbital  period WDMS  binaries that  did not
undergo common  envelope evolution.  This  seems to be confirmed  by a
recent high-resolution imaging campaign  of 90 white dwarfs with known
or   suspected    low-mass   stellar   and    sub-stellar   companions
\citep{farihietal05-1,farihietal10-1}.

The    Sloan   Digital   Sky    Survey   \citep[SDSS,][]{yorketal00-1,
  adelman-mccarthyetal08-1, abazajianetal09-1} has been very efficient
at identifying large  numbers of WDMS binaries \citep{raymondetal03-1,
  silvestrietal07-1,  helleretal09-1},  with   1602  systems  in  Data
Release  (DR)\,6  \citep{rebassa-mansergasetal10-1}.   Intense  radial
velocity studies have  led to the identification of  a large number of
PCEBs     among     the    sample     of     SDSS    WDMS     binaries
\citep{rebassa-mansergasetal07-1,                    schreiberetal08-1,
  schreiberetal10-1}.  A  few examples of  the work already  done with
both the entire  sample of SDSS WDMS binaries,  as well as the
  subset of systems  identified to be PCEBs is  the identification of
many  eclipsing PCEBs,  important for  testing the  models  of stellar
structure   \citep{pyrzasetal09-1,   nebotetal09-1,   parsonsetal10-1,
  parsonsetal11-1,  pyrzasetal11-1}; observational constraints  on the
efficiency   of    the   common   envelope   \citep{zorotovicetal10-1,
  demarcoetal11-1};  strong  evidence   for  a  discontinuity  in  the
strength  of  magnetic  braking  near the  fully  convective  boundary
\citep{schreiberetal10-1}; and  an unambiguous demonstration  that the
majority  of low-mass (He-core)  white dwarfs  are formed  in binaries
\citep{rebassa-mansergasetal11-1}.

Here, we present the final  catalogue of WDMS binaries identified from
SDSS (DR\,7)  spectroscopy, discuss the global properties  of the 2248
systems,  and  describe a  public  interactive  on-line  data base  of
spectroscopic SDSS WDMS binaries.

\section{From SDSS\,I to SDSS\,II}

During  its first  phase of  operations, 2000  to 2005,  SDSS-I mainly
obtained  spectra of galaxies  and quasars  selected from  5700 square
degrees of imaging.  The corresponding data largely dominate the DR\,6
of SDSS  \citep{adelman-mccarthyetal08-1}. From 2005  to 2008, SDSS-II
carried out three distinct surveys:

\begin{itemize}
\item  the Sloan Legacy  Survey, that  completed the  original (mainly
  extragalactic) SDSS\,I imaging and spectroscopic goals;
\item SEGUE, \citep[the SDSS  Extension for Galactic Understanding and
  Exploration,][]{yannyetal09-1} that probed the structure and history
  of the Milky Way, obtaining additional imaging over a large range of
  galactic latitudes as well as spectroscopy for $\sim240\,000$ stars;
\item and the Sloan Supernova  Survey, that carried out repeat imaging
  of the 300 square degree  southern equatorial stripe to discover and
  measure supernovae and other variable objects.
\end{itemize}

The  final data  product  of SDSS\,I  and  most of  SDSS\,II is  DR\,7
\citep{abazajianetal09-1}.   The different design  of SDSS\,II  and in
particular  the inclusion  of a  dedicated target  selection  for WDMS
binaries in SEGUE significantly changes the resulting WDMS binary star
content compared to DR\,6.  Below we review the two main channels that
led to WDMS binary spectra taken by SDSS.

\subsection{WDMS in SDSS\,I}

The main  science driver of  SDSS\,I has been to  acquire spectroscopy
for magnitude-limited samples  of galaxies \citep{straussetal02-1} and
quasars \citep{richardsetal02-1}.  Because of their  composite nature,
WDMS  binaries form  a ``bridge''  in colour  space that  connects the
white     dwarf     locus     to     that    of     low-mass     stars
\citep{smolcicetal04-1}. The blue end  of the bridge, characterised by
WDMS  binaries  with hot  white  dwarf  and/or  late type  companions,
strongly overlaps with the colour  locus of quasars, and was therefore
intensively targeted by SDSS spectroscopy. In contrast, the red end of
the bridge is totally dominated by WDMS binaries containing cool white
dwarfs,  and excluded from  the quasar  program. Some  additional WDMS
binaries were  directly selected  for SDSS spectroscopy  following the
selection      criteria       of      \citet{raymondetal03-1}      and
\citet{silvestrietal06-1}. These criteria are based on the idea that a
WDMS  binary needs  to be  both red  (main sequence)  and  blue (white
dwarf), which is only true for a relatively small fraction of possible
WDMS  binary colours.   In particular,  systems  in which  one of  the
stellar components  dominates the emission are  excluded.  In summary,
the   overall   SDSS\,I   spectroscopic   sample  of   WDMS   binaries
\citep{raymondetal03-1,       silvestrietal07-1,       helleretal09-1,
  rebassa-mansergasetal10-1} is  heavily biased against  WDMS binaries
containing cold  white dwarfs  and/or early-type secondaries,  just as
\citet{schreiber+gaensicke03-1} found for the pre-SDSS sample.

\subsection{WDMS in SDSS\,II}
\label{s-segue}

Within  SDSS\,II, the  Legacy  project is  expected  to identify  WDMS
binaries  with similar  properties, and  at a  similar rate  as within
SDSS\,I. However, SEGUE included  a small number of projects targeting
specific classes of objects, and  we have developed a colour selection
to find WDMS binaries containing either cool white dwarfs and/or early
M-dwarf/late K-dwarf  companions, i.e.  a population  of WDMS binaries
that has  been consistently under-represented  in the SDSS\,I  and all
the previous surveys.

\begin{figure*}
\begin{center}
\includegraphics[angle=270,clip=,width=\linewidth]{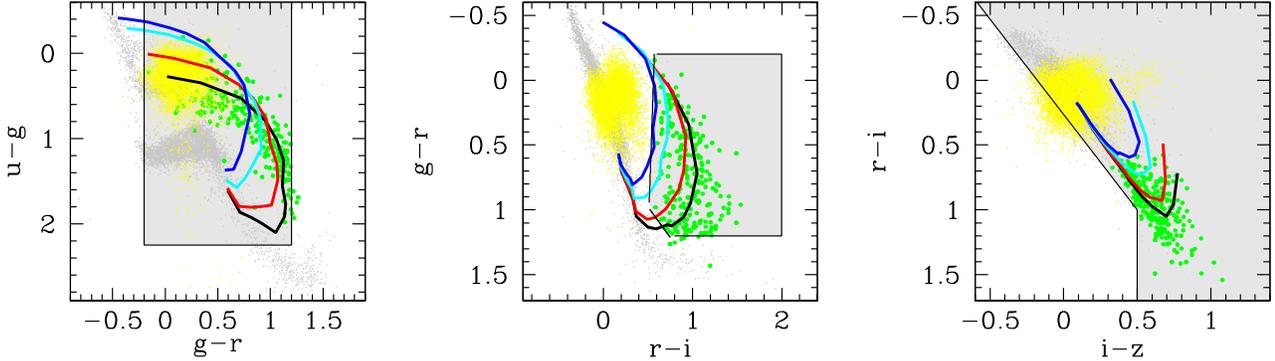}
\end{center}
\caption{\label{g:segcol} Synthetic  WDMS binary colours  in the $u-g$
  vs. $g-r$  (left panel), $g-r$  vs. $r-i$ (centre panel),  and $r-i$
  vs. $i-z$  (right panel) colour-colour diagrams.   Colour tracks are
  shown for WDMS binaries  containing secondary star spectral types K0
  to M6, as  well as white dwarf effective  temperatures of 40\,000\,K
  (blue), 30\,000\,K (cyan), 20\,000\,K (red), and 10\,000\,K (black).
  Colours  for quasars  and single  main sequence  stars are  shown as
  yellow and gray dots respectively.   In gray shaded we represent our
  selection  criteria  specially designed  to  identify WDMS  binaries
  containing  cold  white dwarfs  and/or  early-type companions.   The
  resulting  sample  of  WDMS  binaries identified  by  our  selection
  criteria is represented by green dots.}
\end{figure*}

To that end  we simulated $ugriz$ colours of  WDMS binaries spanning a
broad     range    in     white    dwarf     effective    temperatures
($\Twd=6000-40\,000$\,K)  and companion  star spectral  types (K0--M6)
\citep[see   also][]{schreiberetal07-1}.   Absolute  Johnson   $UBVRI$
magnitudes  for white dwarfs  and for  M/K dwarfs  were taken  from an
updated  version of \citet{bergeronetal95-2}  and \citet{pickles98-1},
respectively.  The  combined  $UBVRI$  magnitudes  were  converted  to
$ugriz$    using    the    empirical   colour    transformations    of
\citet{jordietal06-1}.  Fig.\,\ref{g:segcol}  shows  the loci  of  the
synthetic SDSS  WDMS binary  colours in three  different colour-colour
diagrams ($u-g$  vs.  $g-r$, $g-r$  vs.  $r-i$, and $r-i$  vs.  $i-z$)
and demonstrates  that systems  dominated by the  thermal flux  of the
secondary  star can  be  separated from  quasars  (yellow dots),  main
sequence stars (grey dots), and WDMS binaries dominated by the thermal
flux of the white dwarf by applying the following cuts:

\begin{eqnarray}
\label{eq-segue} 
u-g & < & 2.25  \hspace{0.8cm}g-r  > -19.78(r-i) + 11.13 \nonumber\\
g-r & > & -0.2  \hspace{0.8cm}g-r  <  0.95(r-i) + 0.5\nonumber\\
g-r & < & 1.2   \hspace{0.8cm}i-z  >  0.5  \hspace{0.2cm}\mathrm{for} \hspace{0.2cm}r-i > 1.0\nonumber\\
r-i & > & 0.5   \hspace{0.8cm}i-z  >  0.68(r-i)-0.18 \hspace{0.2cm}\mathrm{for} \hspace{0.2cm}r-i <= 1.0\nonumber\\
r-i & < & 2.0   \hspace{0.8cm}15 <  g  < 20.\nonumber
\end{eqnarray}

These   selection  criteria   (grey-shaded   in  Fig.\,\ref{g:segcol})
optimize the  identification of WDMS binaries consisting  of both cold
white dwarfs and early type secondaries.

\begin{table*}
\caption{The list of 116 plate-pairs and eight single plates with WDMS
  target    selection   that   have    been   observed    in   SEGUE.}
\setlength{\tabcolsep}{0.3ex} \centering
\begin{small}
\begin{tabular}{cccccccccccccc}
\hline
\hline
2303/2318& 2315/2330& 2382/2402& 2397/2417& 2459/2474& 2555/2565& 2667/2671& 2682/2700& 2801/2822& 2854/2869& 2888/2913& 2899/2924& 2909/2934& 2336\\
2304/2319& 2316/2331& 2383/2403& 2441/2443& 2537/2545& 2556/2566& 2668/2672& 2683/2701& 2803/2824& 2855/2870& 2889/2914& 2901/2926& 2910/2935& 2337\\
2305/2320& 2317/2332& 2384/2404& 2442/2444& 2538/2546& 2557/2567& 2669/2673& 2689/2707& 2805/2826& 2856/2871& 2890/2915& 2902/2927& 2911/2936& 2475\\
2306/2321& 2334/2339& 2386/2406& 2445/2460& 2539/2547& 2558/2568& 2670/2674& 2690/2708& 2806/2827& 2857/2872& 2891/2916& 2903/2928& 2938/2943& 2552\\
2307/2322& 2335/2340& 2387/2407& 2446/2461& 2540/2548& 2559/2569& 2676/2694& 2714/2729& 2807/2828& 2858/2873& 2893/2918& 2904/2929& 2939/2944& 2620\\
2308/2323& 2378/2398& 2389/2409& 2447/2462& 2541/2549& 2621/2627& 2677/2695& 2724/2739& 2812/2833& 2859/2874& 2894/2919& 2905/2930& 2940/2945& 2865\\
2310/2325& 2379/2399& 2390/2410& 2449/2464& 2551/2561& 2622/2628& 2678/2696& 2797/2818& 2849/2864& 2861/2876& 2895/2920& 2906/2931& 2941/2946& 2866\\
2312/2327& 2380/2400& 2393/2413& 2452/2467& 2553/2563& 2623/2629& 2680/2698& 2798/2819& 2852/2867& 2862/2877& 2897/2922& 2907/2932& 2963/2965& 2942\\
2313/2328& 2381/2401& 2394/2414& 2457/2472& 2554/2564& 2624/2630& 2681/2699& 2800/2821& 2853/2868& 2887/2912& 2898/2923& 2908/2933&	     &     \\
\hline    
\end{tabular}
\end{small}
\label{t-plates}
\end{table*}

\section{The DR7 SDSS WDMS binary catalogue}
\label{s-finalcat}

In this Section we describe  the final spectroscopic DR\,7 WDMS binary
catalogue.  It  is important to keep  in mind that  the complete DR\,7
catalogue will  be formed by  WDMS binaries identified in  SDSS\,I and
the Legacy survey in SDSS\,II,  and by WDMS binaries identified within
our SEGUE  survey. In what  follows we will  denote the former  as the
SDSS\,I/II WDMS  binary sample,  the latter as  the SEGUE  WDMS binary
sample.

We present first the outcome of the dedicated SEGUE WDMS binary survey
and provide  then a brief review  of the WDMS  binary search algorithm
that  we  apply  to  the  entire SDSS  DR\,7  spectroscopic  data  set
\citep[for  details   see][]{rebassa-mansergasetal10-1}.   Finally  we
estimate the  completeness of the  final SDSS DR\,7  spectroscopy WDMS
binary catalogue.

\subsection{The SEGUE WDMS binary sample}
\label{s-samp}

In October 2005  SEGUE incorporated the colour selection  given in the
previous Section  with the goal of  targeting on average  five, and at
most ten  WDMS binary candidates  per plate-pair.  In addition  to the
colour criteria we requested clean photometry for the selection of our
targets. By the end of SDSS\,II in mid-2008, 116 plate-pairs and eight
single plates including our WDMS binary target selection were observed
in SEGUE (Table\,\ref{t-plates}). However,  during the early stages of
SEGUE,  14  plate-pairs  and  two  single  plates  (those  $<2377$  in
Table\,\ref{t-plates})   considered  de-reddened   $ugriz$  magnitudes
before applying our colour selection,  resulting in a number of single
(unreddened) foreground M-dwarfs being observed.

Using          the           DR\,7          casjobs          interface
\citep{li+thakar08-1}\footnote{http://casjobs.sdss.org/CasJobs/},    we
selected all point-sources within  the footprint defined by the plates
listed  Table\,\ref{t-plates} that have  clean photometry  and satisfy
our  colour selection.   This search  resulted in  10505  unique point
sources, of  which 429 were followed-up  spectroscopically.  Among the
429  spectroscopic  objects,  we  identified  274  as  WDMS  binaries,
corresponding  to   a  ``hit-rate''  of  $\simeq64$   per  cent.   Two
additional     WDMS     binaries     (SDSSJ135643.56--085808.9     and
SDSSJ135930.96--101029.7) were  found on plate 2716 that  has not been
published via DR\,7, and 15 objects were found on the plates done with
de-reddened  magnitudes  that  do  not satisfy  our  colour  criteria.
Cross-matching  these  291  WDMS  binaries with  our  DR\,6  catalogue
\citep{rebassa-mansergasetal10-1}, we  found that 40  systems had SDSS
spectroscopy obtained  as part  of SDSS\,I/II, implying  that $\sim$15
per cent of the targeted  systems are duplicates. Hence, the number of
genuinely new systems identified within SEGUE is 251.

The loci of  the SEGUE WDMS binaries is  shown in Fig.\,\ref{g:segcol}
in  green, a  few systems  outside our  colour selection  (gray shaded
area)   are  those  found   on  the   plates  done   with  de-reddened
magnitudes. WDMS  binaries identified by our SEGUE  survey are flagged
as such in the last column of Table\,6 in the Appendix.

\subsection{The final DR\,7 WDMS binary sample}
\label{s-identwdms}

We searched all new $\sim$0.4 million spectra from SDSS DR\,7 for WDMS
binaries   following   the   template-fitting   method   outlined   in
\citet{rebassa-mansergasetal10-1}. We used as templates 163 previously
identified   WDMS  binaries   from  \citet{rebassa-mansergasetal10-1},
spanning  the whole range  of white  dwarf effective  temperatures and
companion star spectral types,  and calibrated the constraints in both
$\chi^{2}$  and  signal-to-noise ratio  for  each  of  them.  We  then
visually inspected the selected WDMS binary candidates and divided the
systems in  three different categories:  WDMS binary, white  dwarf and
M-dwarf.  Given that  visual inspection of WDMS binaries  in which one
of  the  stellar  components   dominates  the  SDSS  spectrum  can  be
misleading, we complemented the SDSS data with the photometry provided
by  the Galaxy  Evolution  Explorer \citep[$GALEX$;][]{martinetal05-1,
  morrisseyetal05-1}    and   the    UKIRT    Infrared   Sky    Survey
\citep[UKIDSS;][]{dyeetal06-1,    hewettetal06-1,    lawrenceetal07-1,
  warrenetal07-1} and  searched for blue  (red) excess in  the spectra
classified as M-dwarf (white dwarf).  All systems in which we detected
either  a blue  or red  excess  were re-classified  as WDMS  binaries.
Subsequently we inspected the SDSS  images of our selected WDMS binary
candidates, and excluded objects  with morphological problems in their
images.  Finally, we cross-checked our  list of WDMS binaries with the
1602 systems  from DR\,6 \citep{rebassa-mansergasetal10-1}  as well as
the  251 SEGUE  WDMS binaries  identified in  Sect.\,\ref{s-samp}. 390
genuinely new  WDMS binaries were found  in the Legacy  part of DR\,7,
which in addition  to the 251 systems identified  in the SEGUE survey,
bring the total number of new WDMS binaries in DR\,7 to 641.

\subsection{Catalogue completeness}
\label{s-internal}

Here,  we analyse  the completeness  of the  SEGUE WDMS  binary sample
within the  survey footprint defined  by the 240  spectroscopic plates
listed in  Table\,\ref{t-plates}, as well  as the completeness  of the
SDSS\,I/II WDMS binary sample.

We define the completeness of our  SEGUE sample as the number of SEGUE
WDMS    binaries     found    by    our     template-fitting    method
(Sect.\,\ref{s-identwdms}) that have  colours satisfying our selection
criteria (Eq.\,\ref{eq-segue}) divided by  the number of all 274 SEGUE
WDMS binaries found in  Sect.\,\ref{s-samp}. We find a completeness of
100 per cent,  i.e. all SEGUE WDMS binaries  were correctly identified
by our template-fitting method.

\begin{figure}
\centering
\includegraphics[width=\linewidth]{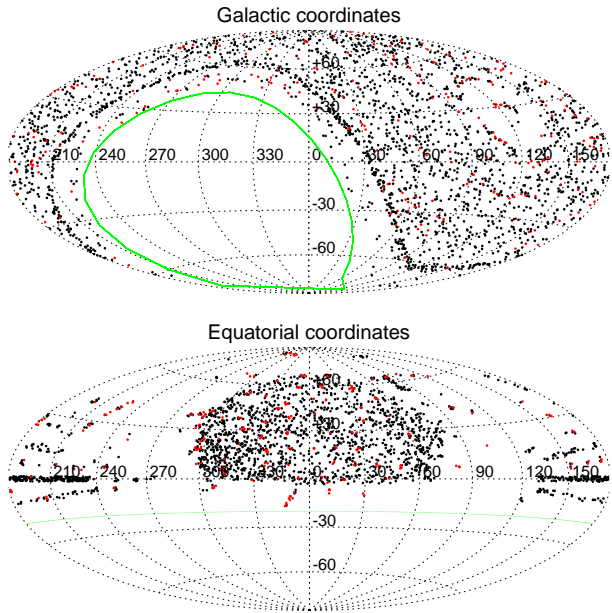}
\caption{Position of SDSS\,I/II (black)  and SEGUE (red) WDMS binaries
  in Galactic and equatorial coordinates.}
\label{f-gal}
\end{figure}

To calculate the completeness of  the SDSS\,I/II WDMS binary sample is
not  straightforward, as it  becomes necessary  to analyse  the entire
Legacy footprint. Such an endeavor is  far away from the scope of this
paper,  however   we  can  estimate  the   completeness  by  analysing
photometric  areas  representative   of  the  SDSS\,I/II  WDMS  binary
population. For this purpose we  used the three colour regions defined
by  \citet{rebassa-mansergasetal10-1}.   For  the  DR\,6  WDMS  binary
catalogue this  resulted in  a completeness of  $\ga98$ per  cent (see
Sect.\,3 in \citealt{rebassa-mansergasetal10-1}).  Here, we identified
only  five   additional  WDMS  binaries  to  the   390  identified  in
Sect.\,\ref{s-identwdms} that  were not found  by our template-fitting
method, implying a completeness of  $\ga98$ per cent. The five systems
are        SDSS\,J001846.79+000237.6,       SDSS\,J020756.15+214027.4,
SDSS\,J133902.65+104136.3,        SDSS\,J150538.90+563353.2,       and
SDSS\,J224122.87+010608.6.   The  spectra of  these  five objects  are
completely dominated  by the  flux of the  secondary star, and  only a
mild blue  excess is  seen in  the blue part  of the  spectra.  Adding
these five  objects to our  SDSS\,I/II sample increases the  number of
new spectroscopic  WDMS binaries to  395.  The \emph{total}  number of
WDMS binaries  in the  entire SDSS DR\,7  thus increases to  2248: 251
form  the SEGUE  WDMS binary  sample,  1997 form  the SDSS\,I/II  WDMS
binary     sample     (1602     within     DR\,6     identified     by
\citealt{rebassa-mansergasetal10-1},    395    identified   in    this
work). Fig.\,\ref{f-gal}  provides the  position of the  complete SDSS
WDMS binary catalogue both in Galactic and equatorial coordinates, and
an  excerpt of  the complete  list  can be  found in  Table\,2 of  the
Appendix.

In  summary, we  conclude that  we  are confident  to have  identified
nearly all  ($\ga98$ per cent)  WDMS binaries within the  entire DR\,7
spectroscopic data  release, and that  our SEGUE survey has  been very
efficient in identifying WDMS binaries  within a colour space that has
so far been neglected.

\begin{figure*}
\begin{center}
\includegraphics[angle=-90,width=0.9\textwidth]{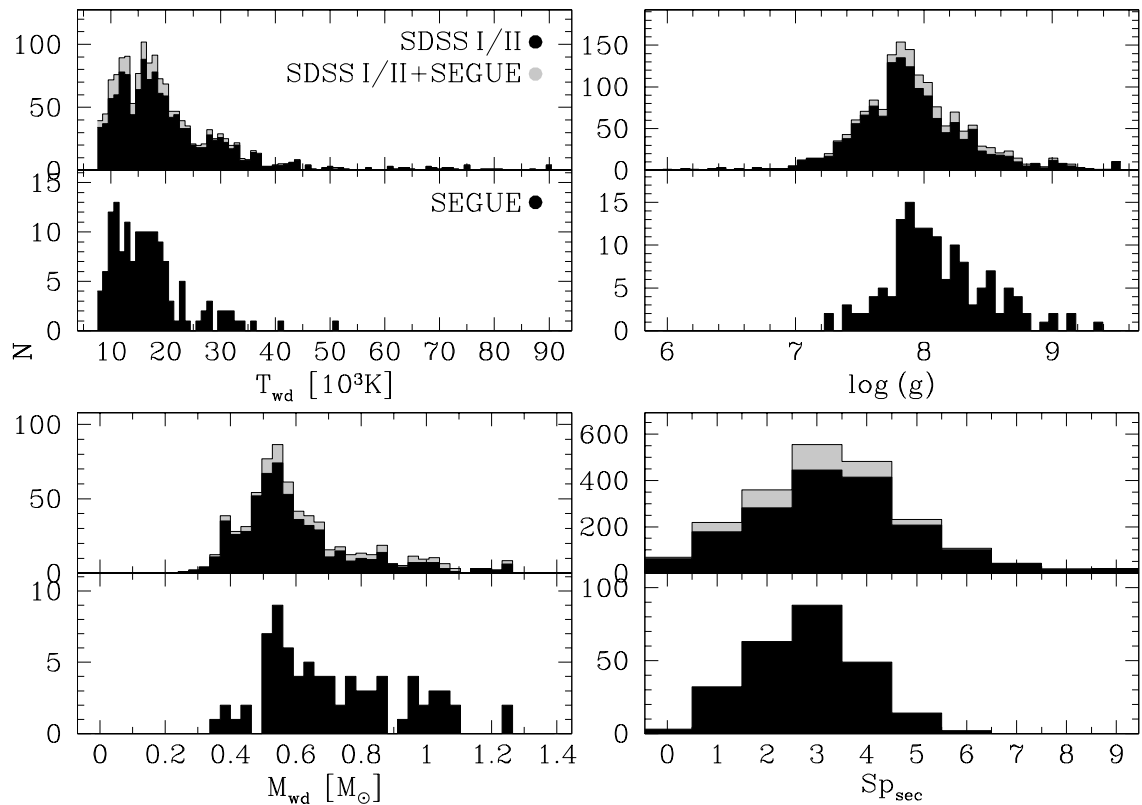}
\caption{Distributions  of  white  dwarf effective  temperatures  (top
  left), surface  gravities (top right) and masses  (bottom left), and
  spectral type  of the companions  (Sp$_\mathrm{sec}$, bottom right).
  Top  panels   show  in  black  and  gray   the  SDSS{\rm{I/II}}  and
  SDSS{\rm{I/II}} plus SEGUE WDMS binary distributions.  Bottom panels
  show in black  the parameter distributions of the  SEGUE WDMS binary
  sample.}
\label{f-histo1}
\end{center}
\end{figure*}

\section{Characterization of the SDSS WDMS binary population}

With more than  2000 systems and being $\ga98$  per cent complete, our
spectroscopic DR\,7  SDSS WDMS binary catalogue represents  the so far
largest and most homogeneous sample  of compact binary stars.  We here
provide   stellar  parameters,   distances,   radial  velocities   and
magnitudes for the complete DR\,7 SDSS WDMS binary catalogue.  Special
attention is  taken to characterise the differences  between the SEGUE
and the SDSS\,I/II WDMS binary samples.

\subsection{Ultraviolet and near-infrared magnitudes}
\label{s-mag}

We  cross-correlated our  list  of WDMS  binaries  with $GALEX$  GR\,6
\citep{martinetal05-1,  morrisseyetal05-1}, which provides  an updated
processing of  the $GALEX$  data compared to  GR\,4, which we  used in
\citet{rebassa-mansergasetal10-1}. Therefore,  the $GALEX$ ultraviolet
magnitudes in this paper supersede  those published in our latest WDMS
binary  catalogue. We  also  cross-correlated all  WDMS binaries  with
UKIDSS   DR\,6  \citep{lawrenceetal07-1,  warrenetal07-1}   to  obtain
near-infrared $yJHK$  magnitudes. Where available,  $GALEX$ and UKIDSS
data are given  in Table\,4 of the Appendix  together with the $ugriz$
SDSS magnitudes.

\subsection{Radial velocities}
\label{s-rvs}

We    measured   radial   velocities    from   the    available   SDSS
sub-spectra\footnote{ Each  SDSS spectrum  is the result  of averaging
  several (typically three) individual exposures, or sub-spectra.}, as
well as from the average SDSS spectra. Radial velocities were measured
fitting two  Gaussians of fixed separation but  free individual widths
and  amplitudes   to  the  \Lines{Na}{I}{8183.27,8194.81},   and/or  a
Gaussian     to      the     H$\alpha$     emission      line     (see
\citealt{rebassa-mansergasetal08-1} for details).  See Table\,3 in the
Appendix for the heliocentric  corrected dates of the observations and
the corresponding radial velocities.

\subsection{Stellar parameters}
\label{s-param}

In order to determine the stellar parameters of our SDSS WDMS binaries
we  used  the  spectral  decomposition/fitting  routine  described  in
\citet{rebassa-mansergasetal07-1,           rebassa-mansergasetal10-1}.
Intensive  tests  of our  white  dwarf  fitting  procedure using  SDSS
spectra  of single white  dwarfs revealed  that we  over-estimated the
errors  on the  white  dwarf parameters  by  a factor  $\sim2$ in  our
previous  SDSS WDMS binary  studies \citep{rebassa-mansergasetal11-1},
we hence  re-fitted the  spectra of all  SDSS WDMS binaries  to obtain
more  realistic  uncertainties of  the  white  dwarf parameters.   

Our fitting  routine follows a  two-step procedure.  Firstly,  a given
SDSS WDMS binary  spectrum is fitted with a  two-component model using
the     M      dwarf     and     white      dwarf     templates     of
\citet{rebassa-mansergasetal07-1}.   We  used  an  evolution  strategy
\citep{rechenberg94-1} to decompose the WDMS binary spectra into their
two individual stellar components.   In brief, this method optimises a
fitness function, in this case a weighted $\chi^2$, and allows an easy
implementation of  additional constraints, such as  e.g. the inclusion
of a  Gaussian fit to the  Balmer emission lines.   From the converged
white  dwarf  plus M-dwarf  template  fit  to  each WDMS  spectrum  we
recorded the spectral type of the  secondary star, as well as the flux
scaling factor between the  M-star template and the observed spectrum.
Subsequently, the best-fit M-dwarf template, scaled by the appropriate
flux scaling  factor, is  subtracted from the  SDSS spectrum,  and the
residual white dwarf spectrum is fitted  with a model grid of DA white
dwarfs  \citep{koesteretal05-1}.  More  specifically, we  fit  the the
normalised  H$\beta$ to  H$\epsilon$  line profiles  to determine  the
white dwarf  effective temperature and surface gravity  ($\log g$). We
exclude H$\alpha$ from this fit as it is in many cases contaminated by
the  flux residuals  of the  M-dwarf.   The equivalent  widths of  the
Balmer lines  go through a  maximum near $\Teff=13\,000$\,K,  with the
exact  value being  a function  of $\log  g$.  Therefore,  $\Teff$ and
$\log g$  determined from Balmer line  profile fits are  subject to an
ambiguity, often referred to  as ``hot'' and ``cold'' solutions.  This
degeneracy  is broken  fitting also  the entire  white  dwarf spectrum
(continuum     plus     lines),     excluding     only     wavelengths
$>7150$\,\AA\ (again to minimise  contamination by flux residuals from
M dwarf  companion). The slope of  the white dwarf  spectrum is mostly
sensitive to $\Teff$, and the  best-fit value from the entire spectrum
is then  used to choose between  the hot and cold  solution.  All fits
are  visually  inspected, and,  where  available,  the  choice of  hot
vs. cold  solution is  further guided by  comparison of  the predicted
ultraviolet  fluxes  to   $GALEX$  measurements.   From  an  empirical
spectral        type-radius        relation        for        M-dwarfs
\citep{rebassa-mansergasetal07-1} and a mass-radius relation for white
dwarfs  \citep{bergeronetal95-2, fontaineetal01-1}  we  calculate then
the radius  of the secondary star and  the mass and the  radius of the
white dwarf respectively.  From the radii and the flux scaling factors
between the WDMS binary components and the white dwarf models and main
sequence star templates, we finally obtain two independent distances.
 
From  the total list  of parameters  we selected  a ``clean''  list by
applying the following restrictions.   As a systematic increase in the
surface  gravity  for  white  dwarfs  below  $\sim$12000\,K  has  been
observed        in       recent       white        dwarf       studies
\citep[e.g.][]{koesteretal09-1}, we  only consider white  dwarf masses
and gravities if  the white dwarf temperature exceeds  this value.  In
order to  avoid contamination  from unreliable stellar  parameters, we
additionally only consider objects with  a relative error in the white
dwarf parameters  of less than  15 per cent.  For  the SDSS{\rm{I/II}}
WDMS binary sample, this resulted  in 1378, 1278 and 579 WDMS binaries
in the  distributions of  white dwarf effective  temperatures, surface
gravities and masses respectively (top panels in Fig.\,\ref{f-histo1},
black).  For the  SEGUE sample the distributions contain  143, 144 and
79 WDMS binaries  respectively (bottom panels in Fig.\,\ref{f-histo1},
black).   The  spectral types  of  the  secondary  stars are  directly
determined   from   the    spectral   template   fitting.    For   the
SDSS{\rm{I/II}}  and   SEGUE  sample,  1768  and   251  WDMS  binaries
containing  M-dwarfs have reliable  spectral types,  respectively, and
are included in the distribution.

\begin{figure}
\begin{center}
\includegraphics[angle=-90,width=\columnwidth]{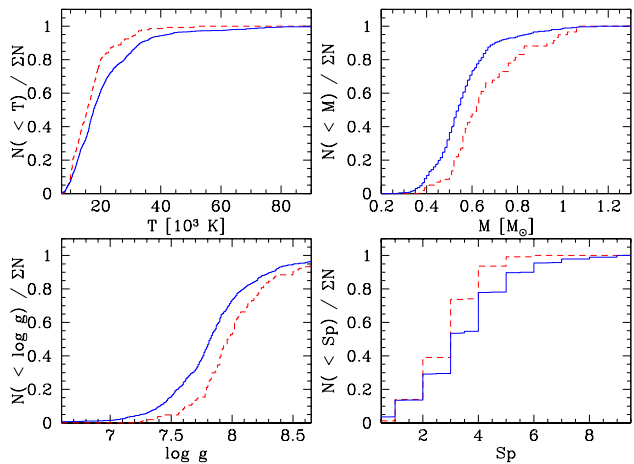}
\caption{White dwarf effective temperature (top left), surface gravity
  (top  right)  and mass  (bottom  left),  and  spectral type  of  the
  companion (Sp$_\mathrm{sec}$, bottom right) cumulative distributions
  obtained  from the  SDSS{\rm{I/II}} sample  of WDMS  binaries (solid
  blue lines),  and those WDMS  binaries identified within  SEGUE (red
  dashed lines).}
\label{f-ks}
\end{center}
\end{figure}

The stellar  parameter distributions of the complete  SDSS WDMS binary
sample  (SDSS{\rm{I/II}} plus  SEGUE) are  shown  in gray  in the  top
panels of Fig.\,\ref{f-histo1} and the cumulative distributions of the
SDSS\,I/II and the SEGUE sample are shown in Fig.\,\ref{f-ks}.

It is  very clear that  the stellar parameters of  the SDSS{\rm{I/II}}
WDMS  binary  sample  differ  from  those in  the  SEGUE  WDMS  binary
population       (black      distributions      Fig.\,\ref{f-histo1}).
Kolmogorov-Smirnov  and  $\chi^2$  (in   case  of  the  spectral  type
distributions)  tests  applied  to  the cumulative  stellar  parameter
distributions in both sub-samples gave probabilities $<10^{-4}$ in all
cases,  clearly indicating  that  the two  populations are  different.
These differences have a  straight forward explanation.  WDMS binaries
detected  within  our   SEGUE  survey  (Sect.\,\ref{s-segue})  are  by
definition of the selection  criteria considerably less blue than WDMS
binaries  from SDSS{\rm{I/II}}.   This favours  the detection  of WDMS
containing cold and massive (and hence small) white dwarfs and early M
spectral types:

\begin{itemize}

\item The upper  left panel of Fig.\,\ref{f-ks} shows  the white dwarf
  effective  temperature  cumulative  distributions.   Inspecting  the
  Figure  it becomes clear  that the  number of  cold white  dwarfs is
  significantly higher in our SEGUE WDMS binary sample.

\item The  upper right and  lower left panels  of Fig.\,\ref{f-histo1}
  and  Fig.\,\ref{f-ks} clearly  show that  the SEGUE  sample contains
  more  systems  with  massive  white  dwarfs with  a  higher  surface
  gravity.

\item The  secondary star spectral type distributions  are provided in
  the bottom right panel of Fig.\,\ref{f-ks} and Fig.\,\ref{f-histo1}.
  As    a   natural    consequence   of    our    selection   criteria
  (Sect.\,\ref{s-segue}), the  sample of SEGUE  WDMS binaries contains
  more secondary stars of spectral types M0-6.

\end{itemize}

The comparison of the SDSS\,I/II and SEGUE WDMS binary samples clearly
demonstrates  that our  selection  criteria (Eq.\,\ref{eq-segue})  was
efficient at  identifying WDMS binaries  containing cold/massive white
dwarfs and early type secondary stars, i.e. a population that has been
underrepresented by all previous surveys.

\subsection{Distances}
\label{s-dist}

Two independent  distances have been  determined for each  system from
the  flux scaling  factors between  the model  (template) fit  and the
observed  fluxes, and  the use  of  a mass-radius  relation for  white
dwarfs and  an empirical  spectral type-radius relation  for M-dwarfs.
The  obtained distances are  compared here.  To avoid  statistical and
systematic  uncertainties in  the comparison  we require  the relative
error in the  white dwarf distances ($d_\mathrm{WD}$) to  be less than
15 per cent,  the effective temperature of the  white dwarfs to exceed
12\,000\,K, and  exclude systems with indications  for both components
being resolved on the SDSS image
\footnote{The flux  contribution of  one or both  stars in  a resolved
  WDMS  binary  spectrum  is  likely  to be  underestimated  and  this
  translates   into  an   underestimated  flux   scaling   factor  and
  overestimated distance.}.  The relative  error in the secondary star
distance  ($d_\mathrm{sec}$)  is  dominated  by  the  scatter  in  the
empirical  M-dwarf spectral  type-radius  relation \citep[][see  their
  Fig.\,7]{rebassa-mansergasetal07-1},  which represents  an intrinsic
uncertainty  rather  than a  statistical  error  in  the fit,  and  we
consequently  do   not  apply  any  error   cut  in  $d_\mathrm{sec}$.
Fig.\,\ref{f-dista}   provides  $d_\mathrm{WD}$   as  a   function  of
$d_\mathrm{sec}$ for the resulting 575 SDSS{\rm{I/II}} (top panel) and
65 SEGUE (bottom panel) WDMS binaries.

An   effect   previously   identified   by   \citet{schreiberetal08-1,
  rebassa-mansergasetal10-1} can be seen in the SDSS{\rm{I/II}} sample
(Fig.\,\ref{f-dista}  top  panel):  for  $27.0\pm2$ per  cent  of  the
systems  the   two  distance  estimates  disagree   at  a  1.5$\sigma$
significance level (red points  in Fig.\,\ref{f-dista}), with the vast
majority     ($21.4\pm2$     per     cent)     of     these     having
$d_\mathrm{sec}>d_\mathrm{WD}$  outliers. Only relatively  few systems
are    found   significantly    below   $d_\mathrm{sec}=d_\mathrm{WD}$
($5.6\pm1$  per  cent).    Conversely,  the  fraction  of  1.5$\sigma$
outliers in  the SEGUE WDMS binary  sample (Fig.\,\ref{f-dista} bottom
panel) not only decreases  but is almost identical above ($10.7\pm3.5$
per      cent)     and      below      ($9.2\pm3.6$     per      cent)
$d_\mathrm{sec}=d_\mathrm{WD}$.

\citet{rebassa-mansergasetal07-1}            interpreted           the
$d_\mathrm{sec}>d_\mathrm{WD}$   outlier  effect  as   resulting  from
magnetic activity  affecting the surface  of the secondary  stars.  If
this was the case, the fraction of active secondary stars in the SEGUE
WDMS binary sample should  be significantly smaller.  One might intend
to interpret  this as being a  consequence of the  SEGUE WDMS binaries
being, on  average, older  (since the white  dwarfs forming  the SEGUE
sample  are systematically  cooler, see  Fig.\,\ref{f-ks}), as  it has
been demonstrated  that older M-dwarfs are  systematically less active
\citep{westetal08-1}.   However, as  we are  considering  only systems
with   white  dwarf   effective   temperatures  exceeding   12\,000\,K
(Fig.\,\ref{f-dista}), this age-effect is expected to be negligible.

\begin{table}
\caption{Percentage of  WDMS binary ages for the  SDSS\,I/II and SEGUE
  samples defined  in Section\,\ref{s-dist}  that are above  and below
  the  activity  lifetimes  estimated  by  \citet{westetal08-1}  as  a
  function    of   spectral    type.}    \setlength{\tabcolsep}{0.8ex}
\centering
\begin{small}
\begin{tabular}{cccccc}
\hline
\hline
Sp. type & Act. lifet. & \multicolumn{2}{c}{SDSS\,I/II} & \multicolumn{2}{c}{SEGUE} \\
         &               & $>$act. lifet & $<$act. lifet& $>$act. lifet & $<$act. lifet \\
         & [Gyr]         & per cent & per cent & per cent & per cent \\
\hline
M0  & 0.8$\pm$ 0.6 & 60 & 40 & -   & -   \\
M1  & 0.4$\pm$ 0.4 & 6  & 94 & 0   & 100 \\
M2  & 1.2$\pm$ 0.4 & 42 & 58 & 37  & 64  \\
M3  & 2.0$\pm$ 0.5 & 57 & 43 & 69  & 31  \\
M4  & 4.5$\pm$ 0.5 & 97 & 3  & 100 & 0   \\
M5  & 7.0$\pm$ 0.5 & 100& 0  & 100 & 0   \\
M6  & 7.0$\pm$ 0.5 & 100& 0  & -   & -   \\
\hline    
\end{tabular}
\end{small}
\label{t-ages}
\end{table}   

\begin{figure}
\begin{center}
\includegraphics[angle=-90,width=\columnwidth]{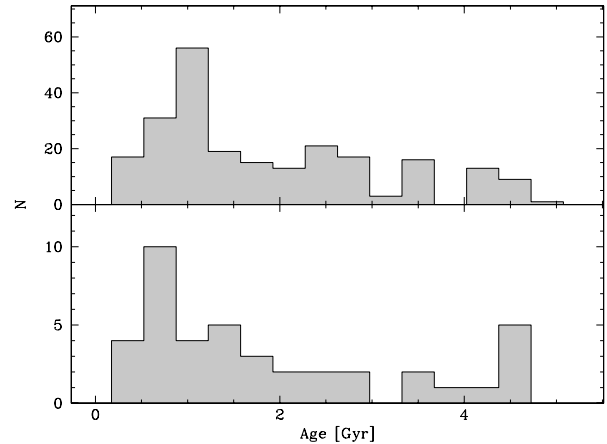}
\caption{\label{f-ages}  Distribution  of   WDMS  binary  ages  for  a
  sub-sample of  231 SDSS{\rm{I/II}} (top panel) and  41 SEGUE (bottom
  panel) systems.  See Section\,\ref{s-dist} for details.}
\label{f-dista}
\end{center}
\end{figure}

To investigate this quantitatively, we  have estimated the ages of the
WDMS   binaries   of   the   SDSS\,I/II  and   SEGUE   subsamples   in
Fig.\,\ref{f-dista}  and compared  them to  the activity  lifetimes of
\citet{westetal08-1}. The total age of each system is given by the sum
of the white  dwarf cooling age and the main  sequence lifetime of the
white dwarf progenitor.  White dwarf cooling ages were calculated from
the  cooling tracks  of \citet{althaus+benvenuto97-1}.   Main sequence
progenitor lifetimes were obtained  from the equations of \citet[][see
  their Appendix]{tuffsetal04-1},  where the main  sequence progenitor
masses  were calculated  using the  initial-to-final mass  relation by
\citet{catalanetal08-1}.   A  relatively   large  percentage  of  WDMS
binaries   are   PCEBs   \citep{schreiberetal10-1},  for   which   the
initial-to-final mass  relation for  single white dwarfs  is generally
not   valid.    However,  \citet[][their   Fig.\,9]{zorotovicetal11-1}
demonstrated  that this  relation is  a good  approximation  for PCEBs
between 0.55  and 0.8$\Msun$, as the  core of the  progenitors of such
white  dwarfs is  almost entirely  developed  at the  onset of  common
envelope evolution.   We therefore considered  only systems containing
white  dwarfs with  masses  in  this range.   The  estimated ages  are
compared  to   the  activity  lifetimes   of  \citet{westetal08-1}  in
Table\,\ref{t-ages}, where we provide  the percentages of systems with
ages  above and  below the  activity lifetime  for  different spectral
types.  For  completeness, the estimated ages are  also illustrated in
Fig.\,\ref{f-ages}.  The  fractions of  systems that should  be active
according  to their age  are very  similar for  WDMS binaries  in both
samples (Table\,\ref{t-ages}),  i.e.  $28\pm3$ per  cent (SDSS\,I) and
$31\pm7$ per cent (SEGUE).  Thus,  the two samples should also contain
approximately   the  same  number   of  $d_\mathrm{sec}>d_\mathrm{WD}$
outliers if those are caused  by activity.  This is apparently not the
case  (Fig.\,\ref{f-dista})  and  we  conclude  that  the  SEGUE  data
questions our previous interpretation of magnetic activity causing the
distance disagreement,  and that this discrepancy  remains an unsolved
issue.          Using         model        spectra         \citep[e.g.
  PHOENIX,][]{hauschildt+baron99-1}  instead of  M-dwarf  templates in
the  decomposition/fitting  of  the   SDSS  WDMS  binary  spectra  and
comparing    both     results    may    provide     useful    insights
\citep[e.g.][]{helleretal09-1}.

\begin{figure}
\begin{center}
\includegraphics[angle=-90,width=\columnwidth]{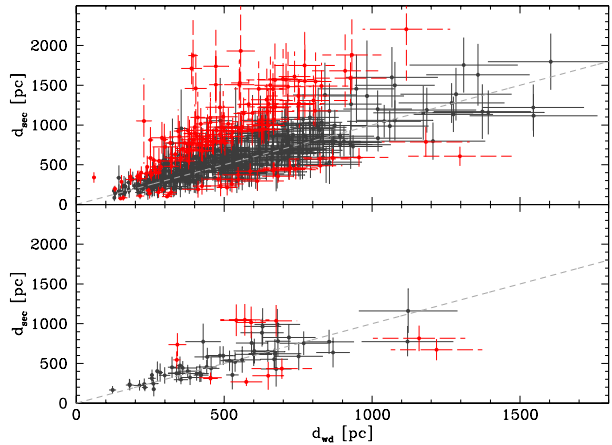}
\caption{Secondary  star  distances  as  a  function  of  white  dwarf
  distances for a sub-sample of 575 SDSS{\rm{I/II}} (top panel) and 65
  SEGUE (bottom panel) WDMS binaries in which the white dwarf distance
  relative  error  is less  than  15  per  cent.  In  red  $1.5\sigma$
  outliers from $d_\mathrm{sec} = d_\mathrm{WD}$ (gray dashed line).}
\label{f-dista}
\end{center}
\end{figure}

\section{The SDSS WDMS binary on-line data base}
\label{s-onlinecat}

We have  presented an updated  version of the spectroscopic  SDSS WDMS
binary  catalogue  incorporating  646   new  systems  from  the  DR\,7
spectroscopic  data  base  (Sect.\,\ref{s-finalcat}).   A  significant
fraction of  the new systems (251)  have been identified  by our SEGUE
WDMS binary survey (Sect.\,\ref{s-segue}).  The entire DR\,7 SDSS WDMS
binary catalogue contains now 2248 objects, each of them associated to
a  large   number  of   stellar  parameters,  radial   velocities  and
photometric   magnitudes.   Given  the   large  amount   of  available
information   we  developed   an  SQL   (Structured   Query  Language)
interactive on-line  data base  for spectroscopic SDSS  WDMS binaries.
This data base  is open for the general  public, however permission to
update the data base will be restricted to our team.

The  data base  allows the  user to  search for  information  in three
different  ways and is  based on  six different  tables that  form the
SQL's db. In  the following we provide instructions on  how to use the
interactive SDSS WDMS binary data base and give details on the content
of each table. More detailed  information and examples can be found on
the web page \emph{http://www.sdss-wdms.org}.

\subsection{Tables description}
\label{s-tables}

Each of  the SDSS  WDMS binaries  in our catalogue  has been  given an
identification number (ID), ranging from 1 to 2248.  This ID is unique
for  each object.   In addition  we have  named our  systems  by their
International     Astronomical      Union     (IAU)     name,     i.e.
SDSSJHHMMSS.SS$\pm$DDMMSS.S, as  well as by an  abbreviated form, i.e.
SDSSJHHMM$\pm$DDMM.   On  the  SQL's  db  the  table  $obj\_iau\_name$
contains the ID and the IAU  name of the 2248 WDMS binaries, the table
$object$  contains  both  the   ID  and  the  abbreviated  names.  The
ultraviolet  $GALEX$ DR\,6,  the  $ugriz$ SDSS  and the  near-infrared
UKIDSS DR\,6 magnitudes plus photometric  errors, as well as the right
ascensions and  declinations of  our targets are  stored in  the table
$magnitudes$.   The  measured   \Ion{Na}{I}  absorption   doublet  and
H$\alpha$  emission   radial  velocities  and   corresponding  errors,
together with the heliocentric corrected dates of the observations can
be found in  the table $rv$.  To these radial  velocities we added the
measurements   presented   in   \citet{schreiberetal10-1}  and   Nebot
Gomez-Moran et al. (2011, submitted).  The stellar parameters together
with the SDSS identifiers (i.e.  modified Julian date, MJD, plate, PLT
and fibre, FIB), the distances, and the WDMS binary spectral types are
given in the table $wdms$. Finally the average stellar parameters (for
several objects multiple SDSS spectra are available) and those derived
from additional follow-up observations performed by us are provided in
the table $mean\_param$. The six tables are linked using the IAU names
and    excerpts   of    each    table   can    be    found   in    the
Appendix\,\ref{s-append}.

\subsection{Queries}

The just described data base can be searched in three different ways.

\begin{itemize}

\item \emph{Search by  object}. This option allows the  user to obtain
  all available information for a specific object or a list of objects
  using either  the IAU or the abbreviated  identifiers.  The provided
  information for each system includes $ugriz$ SDSS, ultraviolet DR\,6
  $GALEX$ and near infra-red DR\,6 UKIDSS magnitudes (when available),
  the  binary and stellar  parameters as  well as  the SDSS  image and
  spectrum, the two  component fits to the SDSS  WDMS binary spectrum,
  and the model  fit to the residual white  dwarf spectrum as obtained
  from our decomposition/fitting routine.

\item \emph{Search by parameter}. The user may define constraints on a
  chosen set  of parameters to obtain  a list of  objects that satisfy
  user defined conditions.  Such constraints can be applied to all the
  stellar  parameters, magnitudes  and radial  velocities, as  well as
  coordinates  and PCEB  orbital periods.   To provide  access  to the
  results of  our follow-up studies, we have  additionally defined the
  following  parameters: (1)  $sigma$,  set  to 3  or  4 depending  on
  whether we detect 3$\sigma$ or 4$\sigma$ radial velocity variations,
  set to 0 when no radial  velocity variations are detected, set to -1
  when  no information  is available  (in  other words  $sigma =  3,4$
  stands for  a PCEB, $sigma =  0$ for a wide  WDMS binary candidate);
  (2) $re$, set to 1 for resolved objects in the SDSS images, set to 0
  for  non-resolved objects;  (3) $spec$,  set to  1 for  objects with
  available follow-up  spectroscopy, set to  0 for objects we  did not
  follow-up; (4) $phot$, set to 1 for objects with available follow-up
  photometry, set to 0 for  objects we did not follow-up; (5) $segue$,
  set  to  1 if  the  considered WDMS  binary  was  identified by  our
  dedicated SEGUE survey, set to 0 otherwise.

\item \emph{Manual  search}. The user may  run his own  SQL scripts on
  the six tables introduced in Sect.\,\ref{s-tables}.  For example, if
  the user wishes to search for WDMS binaries identified as PCEBs with
  at least one of the \Ion{Na}{I} doublet radial velocity measurements
  above  100\,\kms,   no  available  photometric   observations,  SDSS
  $i$-magnitudes of less than 18.5, with declinations between 0 and 30
  degrees, and white dwarf mass  errors less than 0.1\,\Msun, the user
  should type the following:\\
\\
select\\
s.iau\_name, s.mwde, f.i, p.rv\_na\\
from\\
mean\_param as s, rv as p, magnitudes as f\\
where\\
s.iau\_name = p.iau\_name\\
and s.iau\_name = f.iau\_name\\
and p.rv\_na $>$ 100\\
and s.sigma $>$ 0\\
and phot = 0\\
and Mwde between 0.0001 and 0.1\\
and f.i $<$ 18.5\\
and f.decl between 0 and 30\\

\end{itemize}

\section{Summary}
\label{s-concl}

We have identified 646 new  spectroscopic SDSS WDMS binaries; 395 have
been detected  within the spectroscopic SDSS\,I/II  Legacy Survey, the
remaining 251  new discoveries result  from an efficient (64  per cent
success rate) survey  carried out by us within  SEGUE. This survey has
been designed to detect a  population of WDMS binaries containing cold
white dwarfs and/or  early type companion stars, a  population of WDMS
binaries clearly underrepresented in all previous WDMS binary samples.
The  total number  of spectroscopic  SDSS WDMS  binaries  increased to
2248, and we expect this final  DR\,7 SDSS WDMS binary catalogue to be
$\ga$98  per   cent  complete.   Using  an  updated   version  of  our
decomposition/fitting routine  we have determined/updated  the stellar
parameters of  the complete SDSS WDMS binary  catalogue. Comparing the
parameter  distributions  of  the  SDSS\,I/II and  SEGUE  WDMS  binary
samples, we  demonstrated that our  SEGUE survey indeed has  been very
efficient at  identifying WDMS binaries containing  cold white dwarfs.
The  DR\,7  WDMS binary  catalogue  represents  the  largest and  most
homogeneous sample  of compact  binary stars presented  so far  and an
excellent basis for further follow-up studies.  This potential can now
be  explored easily  using our  new user-friendly  interactive on-line
data  base of  SDSS  WDMS binaries  containing  all available  stellar
parameters, radial  velocities and magnitudes,  now publicly available
at \emph{http://www.sdss-wdms.org}.

\section*{Acknowledgments}

ARM and MRS acknowledge financial support from Fondecyt in the form of
grant  numbers  3110049 and  1100782  respectively. ANGM  acknowledges
support by the Centre National d'Etudes Spatial (CNES, ref. 60015). We
thank Ivan Almonacid  for his contributions to the  development of the
SQL data base. We thank Rene Heller for helpful discussions.


\appendix
\section{Excerpts of SQL tables}
\label{s-append}

We provide here excerpts for the six tables on our interactive on-line
data    base.     The    complete    tables     are    available    on
\emph{http://www.sdss-wdms.org}.

\newpage

\begin{table}
\caption{\label{t-object} Table  $object$. Contains the identification
  number  (ID)  and  abbreviated  name for  the  complete  catalogue.}
\setlength{\tabcolsep}{2ex}
\begin{center}
\begin{small}
\begin{tabular}{cc}
\hline
\hline
ID  &  abbreviated name  \\
\hline
0001 & SDSSJ0001+0006 \\
0002 & SDSSJ0004--0020 \\
0003 & SDSSJ0006+0034 \\
0004 & SDSSJ0010+0031 \\
0005 & SDSSJ0012+0010 \\
\hline
\end{tabular}
\end{small}
\end{center}
\end{table}

\begin{table}
\caption{\label{t-objectiau}  Table  $object\_iau\_name$.  Contains  the
  identification  number  (ID)  and international  astronomical  union
  name for the complete catalogue.}  \setlength{\tabcolsep}{2ex}
\begin{center}
\begin{small}
\begin{tabular}{cc}
\hline
\hline
ID  &  IAU name  \\
\hline
0001 & SDSSJ000152.09+000644.7 \\
0002 & SDSSJ000442.00--002011.6 \\
0003 & SDSSJ000611.93+003446.5 \\
0004 & SDSSJ001029.87+003126.2 \\
0005 & SDSSJ001247.18+001048.7 \\
\hline
\end{tabular}
\end{small}
\end{center}
\end{table}


\begin{table*}
\caption{\label{t-rvel} Table $rv$.  Contains the measured \Ion{Na}{I}
  absorption  doublet  and H$\alpha$  emission  radial velocities  and
  corresponding  errors, together with  the heliocentric  Julian dates
  (HJD) of the  observations and the telescope used  for obtaining the
  spectra.   We indicate  by '-'  that no  radial velocity  values are
  available.}  \setlength{\tabcolsep}{2ex}
\begin{center}
\begin{small}
\begin{tabular}{ccccccc}
\hline
\hline
IAU name  & HJD & RV$_\mathrm{Na}$ &error &  RV$_\mathrm{H\alpha}$ & error & telescope \\
          &     &   [$\kms$]       & [$\kms$]& [$\kms$]       & [$\kms$]& \\
\hline
 SDSSJ000152.09+000644.7 &  2451791.8092 &    0.70 &   21.10 &   24.20 &   16.70 & SDSS \\
 SDSSJ001247.18+001048.7 &  2452519.8962 &       - &       - &   12.30 &   18.60 & SDSS \\
 SDSSJ001247.18+001048.7 &  2452518.9219 &  -14.30 &   30.10 &   30.60 &   14.40 & SDSS \\
 SDSSJ001359.39-110838.6 &  2452138.3933 &   28.90 &   16.90 &    0.00 &    0.00 & SDSS \\
 SDSSJ001726.63-002451.1 &  2452559.7852 &  -33.70 &   15.50 &  -30.10 &   11.40 & SDSS \\
\hline
\end{tabular}
\end{small}
\end{center}
\end{table*}

\begin{table*}
\caption{\label{t-magnit}  Table  $magnitudes$.   Contains  the  right
  ascensions  and declinations,  as  well as  the ultraviolet  $GALEX$
  DR\,6,   SDSS  and  near-infrared   UKIDSS  DR\,6   magnitudes  plus
  photometric  errors.  We  indicate  by '-'  that  no magnitudes  are
  available. The errors have not  been included here but are available
  on \emph{http://www.sdss-wdms.org}.}  \setlength{\tabcolsep}{0.8ex}
\begin{center}
\begin{small}
\begin{tabular}{cccccccccccccc}
\hline
\hline
IAU name  & ra    & dec   & $nuv$ & $fuv$ & $u$ & $g$ & $r$ & $i$ & $z$ & $y$ & $J$ & $H$ & $K$ \\
          & [deg] & [deg] &       &       &     &     &     &     &     &     &     &     &     \\
\hline
 SDSSJ000152.09+000644.7  &  0.46704 &  0.11242 &  18.458  &  17.903 &  19.031 &  18.617  &  17.946  &  17.501 &  17.251 &  16.513 &  16.059 &  15.40160 &  15.289  \\
 SDSSJ000442.00--002011.6 &  1.17500 & -0.33656 &   -  &   - &  23.723 &  20.389  &  19.138  &  18.656 &  18.284 &   - &   - &   - &  -  \\
 SDSSJ000611.93+003446.5  &  1.54975 &  0.57958 &  21.780  &   0.000 &  21.383 &  20.926  &  20.129  &  19.005 &  18.381 &  17.536 &  17.056 &  16.58058 &  16.201  \\
 SDSSJ001029.87+003126.2  &  2.62446 &  0.52394 &  20.178  &  19.964 &  21.929 &  20.853  &  19.975  &  19.001 &  18.421 &  17.659 &  17.147 &  16.52821 &  16.363  \\
 SDSSJ001247.18+001048.7  &  3.19658 &  0.18019 &  20.509  &  20.711 &  20.734 &  20.216  &  19.664  &  18.634 &  17.965 &  17.093 &  16.601 &  16.13305 &   0.000  \\
\hline
\end{tabular}
\end{small}
\end{center}
\end{table*}

\begin{table*}
\caption{\label{t-wdmsparam}  Table   $wdms$.   Contains  the  stellar
  parameters  for both  components, the  identifiers for  SDSS spectra
  (MJD,  PLT, FIB),  and  spectral  types of  our  WDMS binaries  (see
  \citealt{rebassa-mansergasetal10-1}   for   a   description).    The
  secondary    star   masses    and   radii    ($M_\mathrm{sec}$   and
  $R_\mathrm{sec}$)  are  obtained   from  the  $Sp-M-R$  relation  by
  \citet{rebassa-mansergasetal07-1}.  We  provide the two  white dwarf
  solutions as  obtained from our  decomposition/fitting routine.  The
  adopted solution  is given by 1  in the first column,  the ruled out
  solution by 0.  The error parameters have not been included here but
  are       accessible       on      \emph{http://www.sdss-wdms.org}.}
\setlength{\tabcolsep}{0.9ex}
\begin{center}
\begin{small}
\begin{tabular}{ccccccccccccccc}
\hline
\hline
sol  & IAU name & type & MJD & PLT & FIB & $\Teff$(wd) & $\log g$ & $\Mwd$ & $\Rwd$ & $d_\mathrm{wd}$ & $Sp$  & $M_\mathrm{sec}$ & $R_\mathrm{sec}$ & $d_\mathrm{sec}$   \\
     &          &      &     &     &     &  [K]        &          &  [$\Msun$]& [$\Rsun]$& [pc]      &      &   [$\Msun$]    &  [$\Rsun$]  &    [pc]   \\ 
\hline

1 & SDSSJ001726.63-002451.1 &  DA/M &  52559 & 1118 & 280 & 15601 &  7.800 &  0.510 &  0.01479 & 504 &  4 & 0.319 & 0.326 &   477 \\ 
0 & SDSSJ001726.63-002451.1 &  DA/M &  52559 & 1118 & 280 & 12681 &  8.060 &  0.640 &  0.01243 & 345 &  4 & 0.319 & 0.326 &   477 \\ 
1 & SDSSJ001726.63-002451.2 &  DA/M &  52518 & 0687 & 153 & 13588 &  8.110 &  0.680 &  0.01203 & 374 &  4 & 0.319 & 0.326 &   503 \\ 
0 & SDSSJ001726.63-002451.2 &  DA/M &  52518 & 0687 & 153 & 15422 &  8.050 &  0.640 &  0.01258 & 418 &  4 & 0.319 & 0.326 &   503 \\ 
1 & SDSSJ001733.59+004030.4 &  DA/M &  51795 & 0389 & 614 & 10918 &  7.180 &  0.270 &  0.02215 & 695 &  4 & 0.319 & 0.326 &   469 \\ 
0 & SDSSJ001733.59+004030.4 &  DA/M &  51795 & 0389 & 614 & 11302 &  6.860 &  0.200 &  0.02726 & 856 &  4 & 0.319 & 0.326 &   469 \\ 
\hline
\end{tabular}
\end{small}
\end{center}
\end{table*}

\begin{table*}
\caption{\label{t-averparam}  Table  $mean\_param$.  Contains  average
  stellar  parameters for  both stellar  components, as  well  as WDMS
  binary spectral types (see \citealt{rebassa-mansergasetal10-1} for a
  description) and  PCEB orbital  periods.  The secondary  star masses
  and radii ($M_\mathrm{sec}$  and $R_\mathrm{sec}$) are obtained from
  the  $Sp-M-R$  relation  by  \citet{rebassa-mansergasetal07-1}.   To
  identify PCEBs and wide WDMS binaries the column $sigma$ is set to 3
  and 4  (representing 3,4$\sigma$ radial velocity  variations), and 0
  respectively, otherwise -1.  To identify resolved objects we set the
  column $re$  to 1,  0 for unresolved  objects.  To  identify objects
  followed-up by our  own team we set $phot$ and $spec$  to 1 if those
  have    been   observed   photometrically    and   spectroscopically
  respectively,  set  to  0  otherwise.   To  identify  WDMS  binaries
  identified   by  our   SEGUE  survey   we  set   $segue$  to   1,  0
  otherwise. Note also that the errors have not been included here but
  are       available       on       \emph{http://www.sdss-wdms.org}.}
\setlength{\tabcolsep}{0.9ex}
\begin{center}
\begin{small}
\begin{tabular}{cccccccccccccccccccc}
\hline
\hline
   IAU name & type & $\Teff$(wd) & $\log g$ & $\Mwd$& $\Rwd$ & $d_\mathrm{wd}$ & $Sp$ &  $M_\mathrm{sec}$ & $R_\mathrm{sec}$ & $d_\mathrm{sec}$ & $sigma$ & $\Porb$ & $phot$ & $spec$ & $re$ & $segue$ \\
            &      &  [K]        &          &  [$\Msun$]& [$\Rsun]$& [pc]     &      &   [$\Msun$]     &  [$\Rsun$]      &    [pc]        &         &  [h]    &        &        &      &          \\ 
\hline
 SDSSJ013335.54+130357.1 &    DA/M & 29052 & 7.955 & 0.665 & 0.01481 & 1626 & 2 & 0.431 & 0.445 & 1252 &    0 &  0 &   0 &    1 &   0 &   0 \\
 SDSSJ013356.07-091535.1 &    DA/M & 12250 & 7.360 & 0.320 & 0.01970 &  573 & 8 & 0.120 & 0.114 &  247 &   -1 &  0 &   0 &    0 &   0 &   0 \\ 
 SDSSJ013418.52+010100.0 &    DA/M & 23343 & 7.710 & 0.490 & 0.01608 & 1022 & 1 & 0.464 & 0.480 & 1146 &   -1 &  0 &   0 &    0 &   1 &   0 \\
 SDSSJ013441.30-092212.7 &    WD/M & 12110 & 7.050 & 0.240 & 0.02446 & 2357 & 2 & 0.431 & 0.445 & 1907 &   -1 &  0 &   0 &    0 &   0 &   0 \\ 
 SDSSJ013504.31-085919.0 &    DA/M &  9401 & 9.060 & 1.230 & 0.00538 &  169 & 4 & 0.319 & 0.326 &  471 &   -1 &  0 &   0 &    0 &   0 &   0 \\
 SDSSJ013716.08+000311.3 &    DA/M & 19193 & 8.390 & 0.860 & 0.00983 &  643 & 2 & 0.431 & 0.445 &  856 &    0 &  0 &   0 &    1 &   0 &   0 \\
\hline
\end{tabular}
\end{small}
\end{center}
\end{table*}


\end{document}